\documentstyle[12pt]{article}
\begin{document}

\title{{\bf Hawking Emission and Black Hole Thermodynamics}
\thanks{Alberta-Thy-22-06, hep-th/0612193,
invited lecture for Session BHT4 of the 11th Marcel Grossmann Meeting
on General Relativity, 2006 July 28}}
\author{
Don N. Page
\thanks{Internet address:
don@phys.ualberta.ca}
\\
Theoretical Physics Institute\\
Department of Physics, University of Alberta\\
Room 238 CEB, 11322 -- 89 Avenue\\
Edmonton, Alberta, Canada T6G 2G7
}
\date{(2006 December 15)}

\maketitle
\large
\baselineskip 18 pt

\section{Introduction}

Black holes are perhaps the most perfectly thermal objects in the
universe, and yet their thermal properties are not fully understood.
They are described very accurately by a small number of macroscopic
parameters (e.g., mass, angular momentum, and charge), but the
microscopic degrees of freedom that lead to their thermal behavior
have not yet been adequately identified.

Strong hints of the thermal properties of black holes came from the
behavior of their macroscopic properties that were formalized in the
(classical) four laws of black hole mechanics \cite{BCH}, which have
analogues in the corresponding four laws of thermodynamics:

The zeroth law of black hole mechanics is that the surface gravity
$\kappa$ of a stationary black hole is constant over its event horizon
\cite{Carter,BCH}.  This is analogous to the zeroth law of
thermodynamics, that the temperature $T$ is constant for a system in
thermal equilibrium.

The first law of black hole mechanics expresses the conservation of
energy by relating the change in the black hole mass $M$ to the
changes in its area $A$, angular momentum $J$, and electric charge $Q$
in the following way:
 \begin{equation}
 \delta M = \frac{1}{8\pi} \kappa \delta A + \Omega \delta J
    + \Phi \delta Q,
 \label{eq:1}
 \end{equation}
where an extended form of the zeroth law implies that not only the
surface gravity $\kappa$, but also the angular velocity $\Omega$ and
the electrostatic potential $\Phi$ are constant over the event horizon
of any stationary black hole.  This first law is essentially the same
as the first law of thermodynamics.

The second law of black hole mechanics is Hawking's area theorem
\cite{Hawkingarea}, that the area $A$ of a black hole horizon cannot
decrease.  This is obviously analogous to the second law of
thermodynamics, that the entropy $S$ of a closed system cannot
decrease.

The third law of black hole mechanics is that the surface gravity
$\kappa$ cannot be reduced to zero by any finite sequence of
operations \cite{Israel}.  This is analogous to the weaker (Nernst)
form of the third law of thermodynamics, that the temperature $T$ of a
system cannot be reduced to absolute zero in a finite number of
operations.  However, the classical third law of black hole mechanics
is not analogous to the stronger (Planck) form of the third law of
thermodynamics, that the entropy of a system goes to zero when the
temperature goes to zero.

Thus the four laws of black hole mechanics are analogous to the four
laws of thermodynamics if one makes an analogy between temperature $T$
and some multiple of the black hole surface gravity $\kappa$, and
between entropy $S$ and some inversely corresponding multiple of the
black hole area $A$.  That is, one might say that $T = \epsilon
\kappa$ and $S = \eta A$, with $8\pi\epsilon\eta = 1$, so that the
$\kappa \delta A/(8\pi)$ term in the first law of black hole mechanics
becomes the heat transfer term $T\delta S$ in the first law of
thermodynamics.

Nevertheless, by a quite independent line of reasoning that was not
directly motivated by Bekenstein's proposal that he had rejected
\cite{BCH}, Hawking made the remarkable discovery that black holes are
not completely black but instead emit radiation \cite{Haw1,Haw2}.
Once he found that the radiation had a thermal spectrum, he realized
that it did make Bekenstein's idea consistent, of a finite black hole
entropy proportional to area, though not Bekenstein's conjectured
value for $\eta$.  In fact, Hawking found that the black hole
temperature was $T = \kappa/(2\pi)$, so $\epsilon = 1/(2\pi)$ and
hence $\eta = 1/4$.  This gives the famous Bekenstein-Hawking formula
for the entropy of a black hole:
 \begin{equation}
 S_{\mathrm{bh}} = S_{\mathrm{BH}} \equiv \frac{1}{4} A.
 \label{eq:2}
 \end{equation}
Here the subscript bh stands for ``black hole,'' and the subscript BH
stands for ``Bekenstein-Hawking.''

\section{Hawking Emission Formulae}

For the Kerr-Newman metrics \cite{Kerr,Newman}, which are the unique
asymptotically flat stationary black holes in Einstein-Maxwell theory
\cite{Israeluniqueness,Carteruniqueness,HE,Robinson,Mazur}, one can
get explicit expressions \cite{Page76} for the area $A$, surface
gravity $\kappa$, angular velocity $\Omega$, and electrostatic
potential $\Phi$ of the black hole horizon in terms of the
macroscopic conserved quantities of the mass $M$, angular momentum $J
\equiv Ma \equiv M^2 a_*$, and charge $Q \equiv MQ_*$ of the hole,
using the value $r_+$ of the radial coordinate $r$ at the event
horizon as an auxiliary parameter:
\begin{eqnarray}
r_+ &=& M+(M^2-a^2-Q^2)^{1/2} = M[1+(1-a_*^2-Q_*^2)^{1/2}],
 \nonumber \\
A &=& 4\pi(r_+^2+a^2) = 4\pi M^2[2-Q_*^2+2(1-a_*^2-Q_*^2)^{1/2}],
 \nonumber \\
\kappa &=& \frac{4\pi(r_+-M)}{A}
 = \frac{1}{2}M^{-1}[1+(1-\frac{1}{2}Q_*^2)(1-a_*^2-Q_*^2)^{-1/2}]^{-1},
 \nonumber \\
\Omega &=& \frac{4\pi a}{A}
 = a_*M^{-1}[2-Q_*^2+2(1-a_*^2-Q_*^2)^{1/2}]^{-1},
 \nonumber \\
\Phi &=& \frac{4\pi Q r_+}{A}
 = Q_* \frac{1+(1-a_*^2-Q_*^2)^{1/2}}{2-Q_*^2+2(1-a_*^2-Q_*^2)^{1/2}}.
 \label{eq:3}
 \end{eqnarray}

Here $a_* = a/M = J/M^2$ and $Q_* = Q/M$ are the dimensionless
angular momentum and charge parameters in geometrical units.  For a
nonrotating uncharged stationary black hole (described by the
Schwarzschild metric), $a_* = Q_* = 0$, so $r_+ = 2M$, $A = 16\pi
M^2$, $\kappa = M/r_+^2 = 1/(4M)$, $\Omega = 0$, and $\Phi = 0$.

Then Hawking's black hole emission calculation \cite{Haw1,Haw2} for
free fields gives the expected number of particles of the $j$th
species with charge $q_j$ emitted in a wave mode labeled by frequency or
energy $\omega$, spheroidal harmonic $l$, axial quantum number or
angular momentum $m$, and polarization or helicity $p$ as
 \begin{equation}
 N_{j\omega lmp} =
  \Gamma_{j\omega lmp}\{\exp[2\pi\kappa^{-1}(\omega-m\Omega-q_j\Phi)]
   \mp 1\}^{-1}.
 \label{eq:4}
 \end{equation}

Here the upper sign (minus above) is for bosons, and the lower sign
(plus above) is for fermions, and $\Gamma_{j\omega lmp}$ is the
absorption probability for an incoming wave of the mode being
considered.

From the mean number $N_{j\omega lmp}$ and the entropy $S_{j\omega
lmp}$ per mode, one can sum and integrate over modes to get the emission
rates of energy, angular momentum (the component parallel to the black
hole spin axis), charge, and entropy by the black hole:
 \begin{equation}
 \frac{dE_{\mathrm{rad}}}{dt} = -\frac{dM}{dt}
 = \frac{1}{2\pi}\sum_{j,l,m,p}\int \omega N_{j\omega lmp} d\omega,
 \label{eq:9}
 \end{equation}
 \begin{equation}
 \frac{dJ_{\mathrm{rad}}}{dt} = -\frac{dJ}{dt}
 = \frac{1}{2\pi}\sum_{j,l,m,p}\int m N_{j\omega lmp} d\omega,
 \label{eq:10}
 \end{equation}
 \begin{equation}
 \frac{dQ_{\mathrm{rad}}}{dt} = -\frac{dQ}{dt}
 = \frac{1}{2\pi}\sum_{j,l,m,p}\int q_j N_{j\omega lmp} d\omega,
 \label{eq:11}
 \end{equation}
 \begin{equation}
 \frac{dS_{\mathrm{rad}}}{dt}
 = \frac{1}{2\pi}\sum_{j,l,m,p}\int S_{j\omega lmp} d\omega.
 \label{eq:12}
 \end{equation}
Here $M$, $J$, and $Q$ (without subscripts) denote the black hole's
energy, angular momentum, and charge.  By the conservation of the
total energy, angular momentum, and charge, the black hole loses these
quantities at the same rates that the radiation gains them.

This is not so for the total entropy, which generically increases. 
The black hole entropy changes at the rate
 \begin{equation}
 \frac{dS_{\mathrm{bh}}}{dt}
 = \frac{1}{2\pi}\sum_{j,l,m,p}\int
 [2\pi\kappa^{-1}(\omega-m\Omega-q_j\Phi)] N_{j\omega lmp} d\omega,
 \label{eq:13}
 \end{equation}
and by using Eq. (\ref{eq:4}), one can show that the total entropy
$S=S_{\mathrm{bh}}+S_{\mathrm{rad}}$ (black hole plus radiation)
changes at the rate
 \begin{equation}
 \frac{dS}{dt}
 = \frac{1}{2\pi}\sum_{j,l,m,p}\int d\omega
 \left[ \pm \ln{(1\pm N_{j\omega lmp})}
   +N_{j\omega lmp}\ln{\left(1+
   \frac{1-\Gamma_{j\omega lmp}}{\Gamma_{j\omega lmp}
     \pm N_{j\omega lmp}} \right)} \right].
 \label{eq:14}
 \end{equation}

For the emission of $n_s$ species of two-polarization massless
particles of spin $s$ from a Schwarzschild black hole
(nonrotating and uncharged) into empty space, numerical calculations
\cite{Page76,Page76b,Page83} gave
 \begin{equation}
 \frac{dE_{\mathrm{rad}}}{dt} = -\frac{dM}{dt}
 = 10^{-5} M^{-2}(8.1830 n_{1/2} + 3.3638 n_1 + 0.3836 n_2),
 \label{eq:15}
 \end{equation}
 \begin{equation}
 \frac{dS_{\mathrm{rad}}}{dt}
 = 10^{-3} M^{-1}(3.3710 n_{1/2} + 1.2684 n_1 + 0.1300 n_2),
 \label{eq:16}
 \end{equation}
 \begin{equation}
 \frac{dS_{\mathrm{bh}}}{dt}
 = - 10^{-3} M^{-1}(2.0566 n_{1/2} + 0.8454 n_1 + 0.0964 n_2).
 \label{eq:17}
 \end{equation}

\section{The Generalized Second Law}

Even if a black hole is not emitting into empty space, there are
strong arguments that the total entropy of the black hole plus its
environment cannot decrease.  This is the {\it Generalized Second
Law} (GSL).  Bekenstein first conjectured it when he proposed that
black holes have finite entropy proportional to their area
\cite{Bek}, and he gave various arguments on its behalf, though it
would have been violated by immersing a black hole in a heat bath of
sufficiently low temperature if the black hole could not emit
radiation \cite{BCH}.

Once Hawking found that black holes radiate \cite{Haw1,Haw2}, he
showed that the GSL held for a black hole immersed in a heat bath of
arbitrary temperature, assuming that the radiation thermalized to the
temperature of the heat bath.  Zurek and Thorne \cite{ZT}, and Thorne,
Zurek, and Price \cite{TZP}, gave more general arguments for the GSL
without this last assumption.  Their arguments were later fleshed out
in a mathematical proof of the GSL for any process involving a
quasistationary semiclassical black hole \cite{FP}.  Other proofs of
the GSL have also been given \cite{Sor1,Sor2,Waldbook,Sor3}.

With some exceptions \cite{Sor2,Sor3}, these proofs so far generally
have two key assumptions: (1) The black hole is assumed to be
quasistationary, changing only slowly during its interaction with an
environment.  It has been conjectured \cite{TZP} that the GSL also
holds, using the Bekenstein-Hawking $A/4$ formula for the black hole
entropy, even for rapid changes in the black hole, but this has not
been rigorously proved.

(2) The semiclassical approximation holds, so that the black hole is
described by a classical metric which responds only to some average or
expectation value of the quantum stress-energy tensor.  This allows
the black hole entropy to be represented by $A/4$ of its classical
horizon.

Other reviews \cite{Wald99,Majumdar,Kiefer02,Israel03,
Jacobson03,Damour,Das,Fursaev,Page2004,Ross} have given more details
of Hawking radiation and black hole thermodynamics.

This research was supported in part by the Natural Sciences and
Engineering Research Council of Canada.

\vfill

\end{document}